\documentclass[fleqn,usenatbib]{mnras}

\usepackage{newtxtext,newtxmath}

\usepackage[T1]{fontenc}

\DeclareRobustCommand{\VAN}[3]{#2}
\let\VANthebibliography\thebibliography
\def\thebibliography{\DeclareRobustCommand{\VAN}[3]{##3}\VANthebibliography}

\usepackage{graphicx}	
\usepackage{amsmath}	
\usepackage{multirow}   
\usepackage{mathtools}  
\usepackage{gensymb}

\AtBeginDocument{\mathcode`v=\varv}

\title[WR\,114 and WR\,142]{Search for particle acceleration in two massive Wolf-Rayet stars using uGMRT observations}

\author[Anindya Saha et al.]{Anindya Saha,\thanks{E-mail: anindya.s1130@gmail.com (A.S)}$^{\star 1}$
Anandmayee Tej,\thanks{E-mail: tej@iist.ac.in (A.T)}$^{1}$
Santiago del Palacio,$^{2}$
Micha\"{e}l De Becker,$^{3}$
Paula Benaglia,$^{4}$
\newauthor
Ishwara Chandra CH,$^{5}$
Prachi Prajapati,$^{6}$
\\
$^{1}$Indian Institute of Space Science and Technology, Thiruvananthapuram 695 547, Kerala, India\\
$^{2}$Department of Space, Earth and Environment, Chalmers University of Technology, Gothenburg, Sweden\\
$^{3}$Space sciences, Technologies and Astrophysics Research (STAR) Institute, University of Li\`{e}ge, Belgium\\
$^{4}$Instituto Argentino de Radioastronom\'{i}a, CONICET-CICPBA-UNLP, Argentina\\
$^{5}$National Centre for Radio Astrophysics, Pune, India\\
$^{6}$Physical Research Laboratory (PRL), Navrangpura, Ahmedabad 380 009, Gujarat, India
}

\date{Accepted XXX. Received YYY; in original form ZZZ}

\pubyear{2023}

\begin{document}
\label{firstpage}
\pagerange{\pageref{firstpage}--\pageref{lastpage}}
\maketitle

\begin{abstract}
Large wind kinetic power of Wolf-Rayet (WR) stars make them ideal targets in low radio frequencies to search for non-thermal emission due to relativistic particle acceleration. In this paper, we present observations of two WR stars, WR\,114 and WR\,142, in Band 4 (550--950 MHz) and Band 5 (1050--1450~MHz) using the upgraded Giant Meterwave Radio Telescope (uGMRT). 
Neither star is detected in the observed frequency bands, nor extended emission associated with them. 
The upper limit to the free-free radio emission from the stellar wind enables us to constrain the mass-loss rate of WR\,114 to $\lesssim \rm 10^{-5}\,M_{ \odot}\,yr^{-1}$; this is a factor three smaller than previously estimated using spectroscopic modelling. If we further assume that the WR stars are binaries, the non-detection of synchrotron emission from the putative wind collision region implies that the stars are either in very wide binary systems away from periastron, or that the stars are in close binary systems with an orbital separation $<70$~AU for WR 114 and $<20$~AU for WR 142. The non-detection of low-frequency radio emission from these two systems thus provides evidence that narrows their nature, though it does not rule them out as bonafide particle-accelerating colliding-wind binaries. 
\end{abstract}

\begin{keywords}
stars: Wolf–Rayet -- stars: WR\,114, WR\,142 -- radio continuum: ISM -- radiation mechanisms: thermal -- radiation mechanisms: non-thermal
\end{keywords}


\section{Introduction}
\label{sec:intro}

Studies of massive stars are of prime importance in stellar astrophysics given their strong mechanical, radiative, and chemical feedback. In particular, their intense radiation field expels and accelerates the outer layers of the star, giving rise to powerful supersonic winds. These winds deposit a large amount of mechanical energy into the interstellar medium (ISM) generating strong adiabatic shocks suitable for relativistic particle acceleration, presumably via diffusive shock acceleration \citep[DSA; e.g.][and references therein]{Drury1983}. Detection of non-thermal (NT) radio emission, which is synchrotron radiation produced by relativistic electrons in the presence of a magnetic field \citep{Blumenthal1970,White1985}, is an observational evidence of particle acceleration. 

About 50 NT radio-emitting binary systems, referred to as particle-accelerating colliding-wind binaries (PACWBs), have been identified\footnote{Updated PACWB catalogue can be found at \url{https://www.astro.uliege.be/~debecker/pacwb/}} \citep{DeBecker2013}. It is not clear how efficient these systems are at converting wind kinetic energy into particle acceleration, but at least for the WR binary Apep, it was recently shown that this fraction can be close to 1~percent \citep{delPalacio2023}, while for WR\,146 it can be as high as 30~percent \citep{Pittard2021}. In an observational breakthrough, NT radio emission from the stellar bubble G2.4+1.4, associated with the presumably single WR star WR\,102, was detected by \cite{Prajapati2019}. In this case, the fraction of the kinetic wind power that is converted into cosmic-ray acceleration was estimated to be of the order of a few percent. Further, NT radio emission has also been detected in bow-shocks of massive runaways stars \citep{Benaglia2010,Moutzouri2022}. For the runaway star BD+43\,3654, \cite{Benaglia2021} estimated that $\sim 10$~percent of the wind kinetic power should be converted into relativistic particle acceleration in the bow shock. 
These observations support the hypothesis that massive stars are relevant sources of Galactic cosmic rays, as estimated by \cite{Seo2018} under the assumption that typically 1--10~percent of their wind luminosity is transferred to relativistic particles.  Moreover, WR stars can also be key to explain the composition of cosmic rays \citep[][and references therein]{Gabici2023}.

The energy budget for NT particle acceleration depends on the wind kinetic power, $P_\mathrm{kin}$. For a star with mass-loss rate $\dot{M}$ and terminal velocity $ v_{\infty}$, it is $P_\mathrm{kin} \approx 0.5\,\dot{M} \, v_{\infty}^2$. In addition, under certain assumptions, the efficiency of particle acceleration scales with the square of the shock (and therefore the wind) velocity \citep{Drury1983}. 
WR stars, which are the evolved counterparts of O-type stars, have high mass-loss rates $\sim (1$--$5) \times 10^{-5}\,\rm M_{ \odot}\,yr^{-1}$ \citep[e.g.,][]{{Abbott1986},{Leitherer1997},{Chapman1999}} and fast winds with velocities ranging between $\sim 700$--$6000\,\rm km\,s^{-1}$ \citep[e.g.,][]{{Nugis1998},{Hamann2000},{Nugis2000}}. Thus, their winds are exceptionally powerful with a typical $P_\mathrm{kin} > 10^{37}\,\rm erg\,s^{-1}$. This amounts to substantial mechanical energy available to drive particle acceleration, thus making this subset of hot and massive stars ideal targets to search for particle acceleration. 

The requirement of large $P_\mathrm{kin}$ sets the primary target selection criteria for this radio investigation of particle acceleration in massive WR stars. Furthermore, there is growing observational evidence of a strong correlation between the detection of NT emission and binarity in O-type and WR systems \citep[e.g.][]{DeBecker2013}. In comparison, detection of NT emission in low radio frequency is reported for only one presumably single non-runaway star system \citep[WR\,102;][]{Prajapati2019}. This detection opens a new window to probe shock physics and assess the role of the local ISM and its interaction with the powerful WR winds as a necessary ingredient for particle acceleration in single massive stars. 

With the aim to explore the above two scenarios, we carried out low-frequency (735 and 1260~MHz) radio observations of WR\,114 and WR\,142, two nearby (distance $\rm \lesssim 2~kpc$) WR stars with large wind kinetic power ($\rm \sim 10^{38}\, erg\,s^{-1}$). Following the discussion in \citet{Benaglia2021}, we derived the tangential velocities of these two stars with respect to its local ISM from the latest proper motion values listed in Gaia DR3 database (see Table~\ref{tab:WRinfo}). The estimated tangential velocities are 10.2 and 18.0 $\rm km\,s^{-1}$ for WR\,114 and WR\,142, respectively, implying that these are not runaway stars \citep[e.g.][]{Eldridge2011}. 
The coordinates and relevant parameters of these stars are listed in Table~\ref{tab:WRinfo}. Located in the Ser OB1 association, WR\,114 is classified as WC5 spectral type. Comparing the strength of emission lines with those in other stars of the same spectral type, \citet{vanderHucht2001} proposed the presence of an OB companion. However, the Potsdam WR spectral modelling was consistent with a single star \citep{Sander2012}. WR\,114 was observed with {\it XMM-Newton} \citep{Oskinova2003}. The non-detection of X-ray emission ($L_{\rm X} \lesssim 2.5 \times 10^{30} \rm erg\,s^{-1}$) was attributed to the high opacity of the metal-rich and dense wind from the WC star. The other target, WR\,142 (spectral type WO2), belongs to the Berkeley 87 cluster and is one of the only four WO stars detected in our Galaxy. 
{\it XMM-Newton} and {\it Chandra} observations reveal weak X-ray emission with excess absorption below 2~keV \citep{Sokal2010}. These authors discussed several possible emission processes, including thermal and non-thermal ones (e.g., inverse Compton scattering) and a possible colliding-wind shock scenario, to interpret their observations, but their results were not conclusive. 

This paper is organized as follows. The observations and data reduction procedure are described in Sect.\,\ref{sec:data}. After reporting on the main results in Sect.\,\ref{sec:results}, we detail in Sect.\,\ref{sec:discussion} the scientific discussion of the two WR stars in the framework of the scenarios introduced above. Our main conclusions are finally provided in Sect.\,\ref{sec:conclusion}.

\begin{table}
\begin{center}
\caption{Parameters of WR\,114 and WR\,142.}
\begin{tabular}{ l  c  c}
\hline\hline %
 & WR\,114 & WR\,142 \\
 \hline
RA (J2000) & $\rm 18^h23^m16.3^s$ & $\rm 20^h21^m44.3^s$ \\
DEC (J2000) & $-13\degree 43'26.1''$ & $+37\degree 22'30.5''$ \\
Spectral type$^{a}$ & WC5 & WO2\\
Distance ($D$; kpc)$^{b}$ & 1.97$\pm 0.09$ & 1.68$\pm 0.04$ \\
$\mu_{\alpha}$ ($\rm mas\,yr^{-1}$)$^b$ & $0.12\pm0.03$ & $-3.44\pm0.01$ \\
$\mu_{\delta}$ ($\rm mas\,yr^{-1}$)$^b$ & $-1.99\pm0.02$ & $-6.37\pm0.01$ \\
$T_{*}$ (kK)$^{c}$ &  79 & 200 \\
$R_{*}$ ($\rm R_{\odot}$)$^{c}$ & 2.68 & 0.80 \\
$\dot{M}$ ($\rm M_{ \odot}\,yr^{-1}$)$^{c}$ & $\rm 3.1\times10^{-5}$ & $\rm 1.6\times10^{-5}$ \\
$v_{\infty}\,\rm(km\,s^{-1})$$^{c}$ & 3200 & 5000 \\
$P_\mathrm{kin}$ (erg\,$\rm s^{-1}$) &$\rm 1.0\times10^{38}$ & $\rm 1.3\times10^{38}$ \\
\hline
\end{tabular}
\label{tab:WRinfo}
\begin{flushleft}
{\bf Note:} $^{a}$\citet{Smith1968} for WR\,114, \citet{Barlow1982} and \citet{Kingsburgh1995} for WR\,142; $^{b}$\textit{Gaia} DR3 data \citep{GaiaDR32022}; $^{c}$\citet{Sander2019}.
\end{flushleft}
\end{center}
\end{table}
%
\section{Observations and data reduction}
\label{sec:data}
%
\begin{table*}
\begin{center}
\caption{Details of observations and the obtained radio maps.}
\begin{tabular}{ l  c  c  c  c  c}
\hline\hline %
\multirow{2}{*}{} &\multicolumn{2}{c}{WR\,114} & \multicolumn{2}{c}{WR\,142} \\
& Band~4 & Band~5 & Band~4 & Band~5 \\
& (550--950 MHz) & (1050--1450 MHz) & (550--950 MHz)& (1050--1450 MHz)\\
\hline
Observation date  & 27/08/2020 & 26/06/2020 & 24/06/2020 & 22/06/2020\\
Flux calibrator & 3C48, 3C286 & 3C48, 3C286& 3C48 & 3C48\\
Phase calibrator  & 1822$-$096 & 1911$-$201 & 2052+365 & 2052+365\\
Angular resolution  & 4.3\arcsec$\times$3.4\arcsec & 2.7\arcsec$\times$1.6\arcsec & 4.1\arcsec$\times$3.3\arcsec & 2.3\arcsec$\times$1.8\arcsec\\
\textit{rms} ($\mu$Jy\,beam$^{-1}$) &  41 & 22 & 37 & 32\\
\hline
\end{tabular}
\label{tab:dataobs}
\end{center}
\end{table*}
%
We probed the radio continuum emission associated with WR\,114 and WR\,142 using the uGMRT situated in Pune, India. 
GMRT offers a Y-shaped array configuration of 30 fully steerable antennas of 45-m diameter. 
The central square km houses 12 randomly placed antennas and the remaining 18 antennas are uniformly distributed in the arms, with 6 on each arm.
With this hybrid array arrangement, GMRT provides largest and smallest baselines of 25~km and 100~m, respectively, which makes it capable of probing radio emission at both small and large spatial scales. The largest detectable structures in Band 4 and 5 are 17 and 7~arcmin, respectively\footnote{GMRT Observer’s Manual (\url{http://www.ncra.tifr.res.in/ncra/gmrt/gmrt-users/observing-help/manual_7jul15.pdf)}}. Details of the GMRT system can be found in \citet{Swarup1991} and \citet{Gupta2017}.

Dedicated observations (Project code: 38\_070) were carried out for both sources in Band 4 (550--950 MHz) and Band 5 (1050--1450~MHz) with the GMRT Wideband Backend (GWB) correlator configured to have a bandwidth of 400~MHz across 4096 channels.
Primary calibrators were observed at the beginning and end of the observation for flux and bandpass calibration.
The phase calibrators were observed after each scan (30~mins) of the target to calibrate the phase and amplitude variations over the entire observing period. Details of the observations are given in Table \ref{tab:dataobs}.

The data reduction process of flagging, calibration, imaging, and self-calibration was done using the \texttt{CAPTURE}\footnote{\url{https://github.com/ruta-k/CAPTURE-CASA6.git}} continuum imaging pipeline for uGMRT \citep{KaleIshwaraChandra2021}, which utilizes tasks from Common Astronomy Software Applications \citep[\texttt{CASA},][]{McMullin2007}.
The \citet{PerleyButler2017} scale was employed to set the flux density calibration.
After the initial rounds of editing and calibration, we used the multi-term multi-frequency synthesis \citep[MT-MFS; see][]{RauCornwell2011} algorithm in the \textit{tclean} task to account for possible deconvolution errors in wide-band imaging. In the pipeline, five rounds of phase-only self-calibration were performed before making the final image.
These maps were then corrected for primary beam gain. All images used in our analysis are primary beam corrected.


\section{Results} \label{sec:results}
\begin{figure*}
    \centering
    \includegraphics[width=\linewidth]{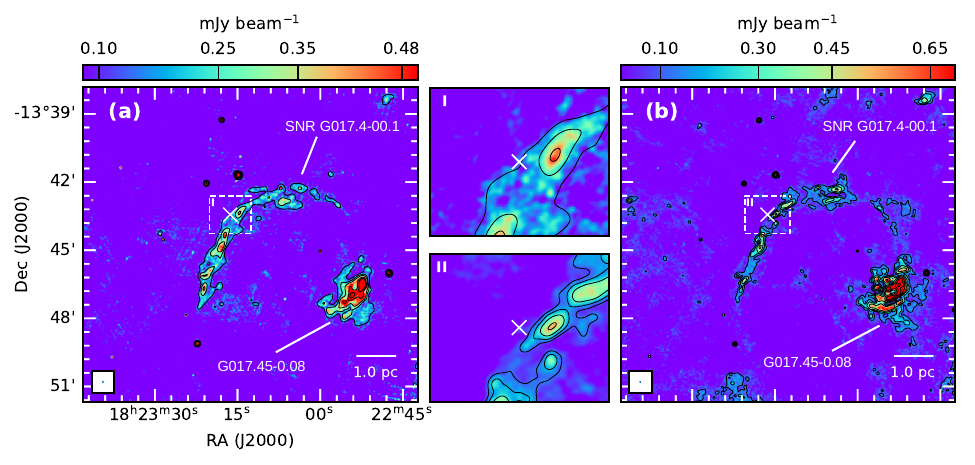}
    \caption{Radio maps of the region around WR\,114, obtained using uGMRT GWB data. Both maps are convolved to a circular beam of size 5.0~arcsec. The white cross ($\times$) in all panels shows the position of WR\,114.
    (a) Map at Band 4 (550$-$950 MHz). The colorscale indicates the flux density. The black contours overlaid correspond to the levels of 
    0.1, 0.2, 0.3, 0.5, 0.7 mJy\,beam$^{-1}$.
    Zoomed in view of the region closer to the WR star (marked I) is also shown.
    (b) Same but for Band 5 (1050--1450 MHz). The contours correspond to the levels of 0.1, 0.2 , 0.3, 0.5, 0.7, 1.0, 1.4 mJy\,beam$^{-1}$. 
    In all panels, the contours are smoothed over 5 pixels using a Gaussian kernel. The locations of the SNR and H\,{\small II} region are marked. Zoomed in view of the region closer to the WR star (marked II) is also shown.}
    \label{fig:WR114radio}
\end{figure*}

\begin{figure*}
     \centering
    \includegraphics[width=\linewidth]{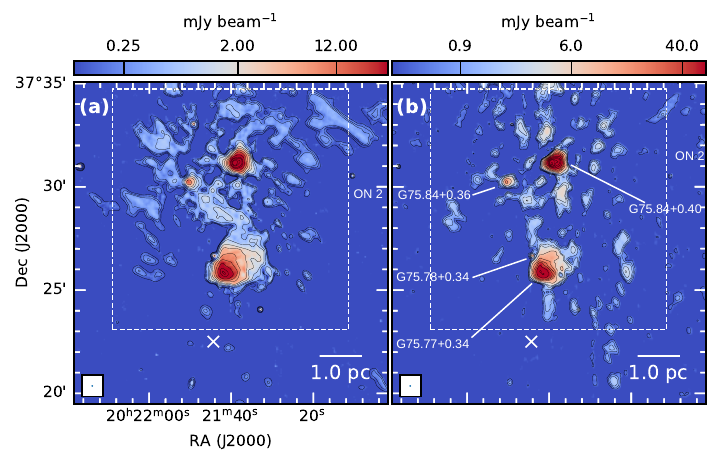}
    \caption{Radio maps of the region around  WR\,142 obtained using uGMRT GWB data. Both maps are convolved to a circular beam of size 5.0~arcsec. For both panels, the white cross ($\times$) marks the position of WR\,142. (a) Map at Band 4 with contour levels of 0.1, 0.2, 0.3, 0.5, 0.9, 1.8, 3.2, 7.0, 14.0, 28.0, 49.0, 57.8, 70.0, 73.5 mJy\,beam$^{-1}$. 
    (b) The same but for Band 5 with contour levels of 0.4, 0.7, 1.2, 2.2, 3.0, 7.2, 15.6, 24.7, 36.4, 65.0, 114.4, 149.5, 184.6, 208.0, 245.7 mJy\,beam$^{-1}$. 
    For both the panels, the contours are smoothed over 5 pixels using a Gaussian kernel.
    The box shows the approximate extent of the massive star forming region ON 2 with the positions of the identified H\,{\small II} regions marked.}
    \label{fig:WR142radio}
\end{figure*}

The radio maps of WR\,114 and WR\,142 are shown in Figures~\ref{fig:WR114radio} and \ref{fig:WR142radio}, respectively. In these maps the positions of the WR stars and other known radio sources are marked. The details of the maps (resolution and {\it rms}) are listed in Table \ref{tab:dataobs}. The maps presented in the figures have been convolved with a beam size of 5.0~arcsec.

There is no radio emission detected for these stars in the observed frequency bands. Of particular mention is the field of WR\,114: The Band 4 map shows faint, diffuse emission at the location of the WR star. However, this emission is most likely part of an arc-like structure associated with the supernova remnant (SNR) SNR G017.4$-$00.1 \citep{Brogan2006,Green2009} in the background along the line-of-sight toward WR\,114, at an estimated distance of 18.6 kpc \citep{Pavlovic2013}.
Other radio sources detected in both fields are marked in the respective figures. 
Thus, these maps give only upper limits to the flux density from the stars. From the achieved {\it rms} level of the uGMRT maps, for WR\,114 we derive $3\sigma$ upper limits of 123~$\mu$Jy ($\nu L_{\nu} < 4.2 \times 10^{26} ~\rm{erg\,s^{-1}}$) and 66~$\mu$Jy ($\nu L_{\nu} < 3.9 \times 10^{26} ~\rm{erg\,s^{-1}}$) at 735 and 1260~MHz, respectively. For WR\,142, the upper limits are 111~$\mu$Jy ($\nu L_{\nu} < 2.8 \times 10^{26} ~\rm{erg\,s^{-1}}$) and 96~$\mu$Jy ($\nu L_{\nu} < 4.1 \times 10^{26} ~\rm{erg\,s^{-1}}$) at 735 and 1260~MHz, respectively.

Both WR stars have been previously observed at higher frequencies. 
In their search for radio emission towards WR stars, \citet{Chapman1999} presented ATCA observations of WR\,114. No emission was detected in none of the observed frequency bands.
Both WR stars are also undetected in a VLA 6~cm observation presented by \citet{Abbott1986}.  In addition, \citet{Cappa2004} observed WR\,114 and WR\,142 at 3.6~cm as part of a VLA survey of Galactic WR stars. In their study, WR\,114 is classified as a probable detection (with signal-to-noise ratio of $\approx$4), whereas WR\,142 is listed as non-detection. It is to be noted that these authors advocate further confirmation of probable detections. Additionally, given the beam size ($\rm 9\,arcsec \times 6\,arcsec$) of these observations, there could be contribution from the background SNR. 
The 3$\sigma$ upper limits from the above studies along with the ones obtained in this work are compiled in Table~\ref{tab:flux_wr114+142}.
%
\begin{table}
\begin{center}
\caption{Flux density measurements of WR\,114 and WR\,142.}
\begin{tabular}{c c c c c }
 \hline
 \hline
 Wavelength & Frequency & Resolution & $S_\nu$ & Ref\\
 (cm) & (GHz) & (\arcsec $\times$ \arcsec) & (mJy) & \\
  \hline
 &&  WR\,114 &&\\
  \hline
 3   & 10 & 1$\times$1      & $<$0.45 & 1\\
 3.6 & 8.3 & 9$\times$6      & 0.15    & 2\\
 \multirow{ 2}{*}{6}   &  \multirow{ 2}{*}{5} & 2$\times$2      & $<$0.45 & 1\\
     &  & 3.5$\times$3.5  & $<$0.3  & 3\\
 13  & 2.3 & 8$\times$5      & $<$0.54 & 1\\
 20  & 1.5 & 12$\times$8     & $<$1.17 & 1\\
23.8 & 1.26 & 2.7$\times$1.6     & $<$0.066 & This work\\
40.8 & 0.735 & 4.3$\times$3.4     & $<$0.123 & This work\\
  \hline
 && WR\,142 &&\\
  \hline
 3.6  & 8.3 & 9$\times$6     & $<$0.9 & 2\\
 6    & 5 & 1.2$\times$1.2 & $<$0.6  & 3\\
 23.8 & 1.26 & 2.3$\times$1.8     & $<$0.096 & This work\\
 40.8 & 0.735 & 4.1$\times$3.3     & $<$0.111 & This work\\
  \hline
 \end{tabular}
\label{tab:flux_wr114+142}
\end{center}
{\bf References.} (1) \citet{Chapman1999}; (2) \citet{Cappa2004}; (3) \citet{Abbott1986}.
\end{table}
%
\begin{figure*}
\includegraphics[width=0.49\linewidth]{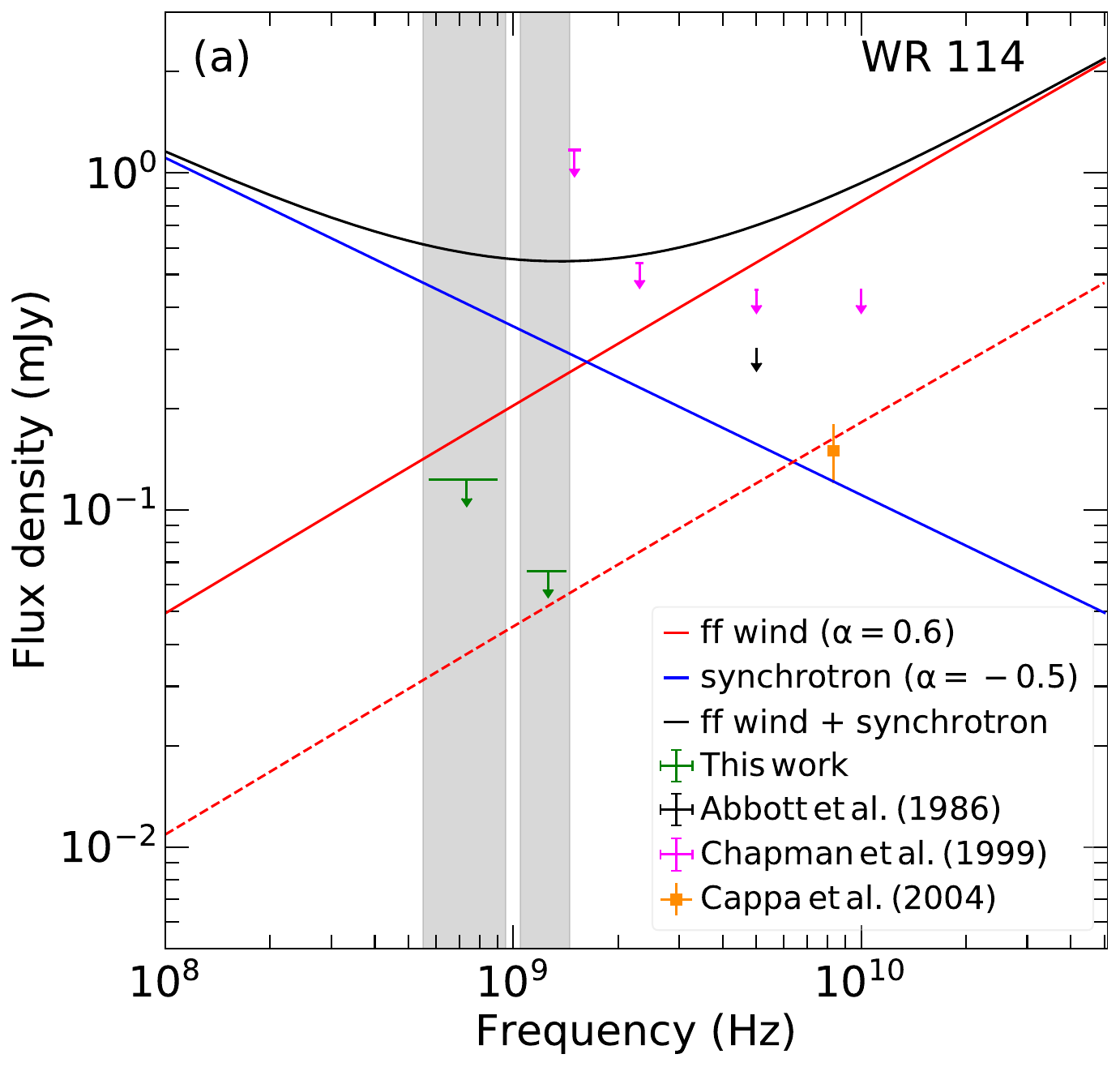}
\includegraphics[width=0.49\linewidth]{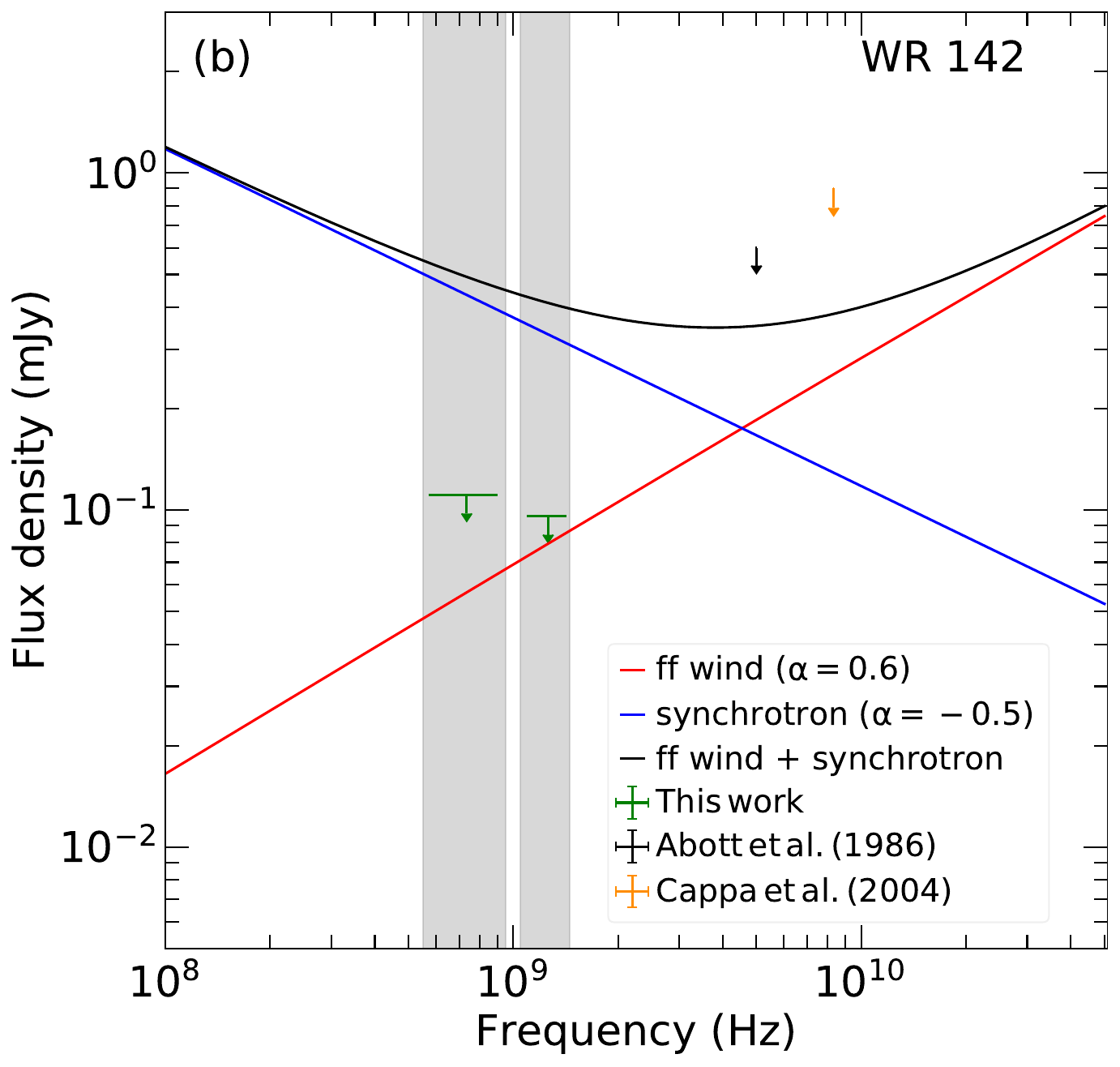}
\caption{Expected spectral energy distribution of WR\,114 (left) and WR\,142 (right) at radio frequencies. The red solid line represents the free-free emission from the stellar wind estimated using Eq. \ref{equ:S_nu} and the parameters listed in Table \ref{tab:WRinfo}. 
The blue solid line represents the lower boundary of the predicted flux density for NT emission in CWB system. The combination of these two is shown as black curve. For WR\,114, the red dashed line represents the 
upper boundary of the free-free flux density estimated using $\dot{M}\,\lesssim$ $\rm 10^{-5}\,M_{ \odot}\,yr^{-1}$ (see text in Sect. \ref{sec:results-singlestellarsys}). The 3$\sigma$ upper limits and the flux density values are shown with markers. The shaded regions represent the bands of our radio observations.}
\label{fig:flux_predict}
\end{figure*}
%
\section{Discussion} \label{sec:discussion}
\subsection{Size of the stellar wind photosphere} \label{sec:photosphere}
The stellar winds consist of ionized gas with a density profile that decreases with the distance to the star as $\propto r^{-2}$. This material can be opaque to the propagation of low-frequency photons due to free-free absorption (FFA). Thus, determining the size of the wind photosphere is crucial for interpreting the radio observations. 

We define the effective radius $R_{\rm \nu}$ that corresponds to an optical depth of unity at a frequency $\nu$. This determines the minimum size from which radiation observed at a frequency $\nu$ can originate, as beyond $R_\nu$ the wind material is optically thin and contributes to the total emission received. 
The characteristic radius of free-free emission region for a spherically symmetric wind of a star with a uniform mass-loss rate $\dot{M}$ and terminal velocity $v_\infty$ can be expressed as \citep{Wright1975,Daley2016} 
\begin{align}
    \left[\frac{R_{\nu}}{\rm cm} \right] &= 1.75\times 10^{28} {\gamma}^{1/3} Z^{2/3} g^{2/3}_\mathrm{ff} T^{-1/2}_\mathrm{e}
    \nonumber \\
    &\times \left( \left[\frac{\dot{M}}{\rm M_{ \odot}\,yr^{-1}} \right] \left[\frac{\rm km\,s^{-1}}{v_\infty}\right] \frac{ 1}{\sqrt{f_\mathrm{w}}\,\mu\nu}\right)^{2/3}, 
    \label{equ:R_nu}
\end{align}
where the free-free Gaunt factor ($g_\mathrm{ff}$) can be approximated as \citep{Leitherer1991}
\begin{equation}
    g_\mathrm{ff} = 9.77\,\left( 1 + 0.13\log \frac{T^{3/2}_\mathrm{e}}{Z \nu} \right).
   \label{equ:gff}
\end{equation}
In these equations, $f_\mathrm{w}$ is the volume filling factor which takes into account the stellar wind clumping \citep{Puls2008}; $T_\mathrm{e}$ is the electron temperature of the wind;   
$\mu$, $Z$, and $\gamma$ are the mean molecular weight, the {\it rms} ionic charge, and the mean number of electrons per ion, respectively. 
Taking the values of $\dot{M}$, and $v_{\infty}$ from Table~\ref{tab:WRinfo}, and considering $T_\mathrm{e} \approx 0.3\,T_{*}$ at radius much greater that $R_{*}$ \citep[see][]{Drew1990}, $f_\mathrm{w} = 0.2$, $\mu = 4.0$, 
$Z = 1.005$, and $\gamma = 1.01$ \citep{Leitherer1995}, we calculate the effective radii of the stellar wind photosphere for WR\,114 and WR\,142 as a function of frequency. 
At the uGMRT frequencies of 735 and 1260~MHz (the effective frequencies of Bands 4 and 5, respectively), $R_{\nu}$($\tau_\mathrm{ff} = 1$) is estimated to be 104 and 71~AU, respectively for WR\,114, and 34 and 23~AU, respectively for WR\,142.
Taking the distance estimates, these translate to angular sizes of $0.05$ and $0.04$~arcsec for WR\,114, and $0.02$ and $0.014$~arcsec for WR\,142, respectively. Comparing with the resolution of the maps (see Table \ref{tab:dataobs}), the stellar wind photosphere would be unresolved in the uGMRT maps, and thus the emission from the stellar winds would be point-like. In addition, we can expect that any emission from the direct surroundings of the WR stars within these photospheric radii will be severely free-free absorbed.

\subsection{Interpreting radio non-detection}
Over the last few decades, investigation of large samples of systems containing OB and WR stars 
\citep[e.g.,][]{Abbott1986,Leitherer1995,Leitherer1997,Chapman1999,Cappa2004,Montes2009,DeBecker2013} and studies of individual targets \citep[e.g.,][]{Williams1997,Dougherty2005,Oconnor2005,Benaglia2005,Benaglia2010,Benaglia2015,Benaglia2019,Prajapati2019,DeBecker2019}
have revealed the presence of both thermal and NT radio emission. 
WR stars are often seen to be in binary (or higher multiplicity) systems 
\citep[e.g.,][]{Meyer2020}. Hence, for the two WR stars studied here, even though there is no conclusive observational evidence of binarity, we cannot rule out this possibility.  
Consequently, our interpretation of the results obtained from radio observations will consider both scenarios.

\subsubsection{Single stellar systems}
\label{sec:results-singlestellarsys} 
For single WR stellar systems, in principle, the observed radio emission is expected to be due to thermal free-free radiation from the partially optically thick stellar wind with a canonical spectral index $\alpha=0.6$ \citep{{PanagiaFelli1975,Wright1975}}.
The flux density due to free-free emission ($S_{\nu}^{\rm ff}$) in the ionized stellar wind of a star with a uniform mass-loss rate, and isothermal outflow with constant wind velocity can be written as \citep{PanagiaFelli1975,Wright1975}
\begin{eqnarray}
        \left[\frac{S_{\nu}^{\rm ff}}{\rm Jy} \right] = 23.2\,\left( \left[\frac{\dot{M}}{\rm M_{ \odot}\,yr^{-1}} \right] \left[\frac{\rm km\,s^{-1}}{v_\infty} \right] \frac{ 1}{\sqrt{f_\mathrm{w}}\,\mu}\right)^{4/3} 
        \nonumber \\
        \times \left( \gamma g_\mathrm{ff}Z^{2}\nu\, \left[\frac{\rm kpc}{D} \right]^3 \right)^{2/3},
    \label{equ:S_nu}
\end{eqnarray}
where $D$ is the distance to the source,
and the rest of the parameters have the same meaning as in Eq.~\ref{equ:R_nu}. Figures \ref{fig:flux_predict}(a) and (b) show the predicted radio spectral energy distribution (SED) for the stellar wind of WR\,114 and WR\,142, respectively. The estimated uGMRT upper limits together with those from the literature (see Table \ref{tab:flux_wr114+142}) are also plotted in the figures.

For WR\,114, considering the predicted
flux density values (169.1 $\mu$Jy at Band 4 and 235.1 $\mu$Jy at Band 5), there should have been detections at $\sim 4\sigma$ and $\sim 10\sigma$ level in uGMRT Bands 4 and 5, respectively.
Similar inferences can be drawn for four of the data points taken from \citet{Abbott1986,Chapman1999,Cappa2004}. 
Hence, the non-detection strongly suggests that either some of the adopted stellar and/or wind parameters are not accurate, or that the assumed stellar wind model is not appropriate for this WR system.
The predicted flux density is most sensitive to the mass-loss rate (Eq.~\ref{equ:S_nu}). 
For WR\,114, the uGMRT upper limit in Band 5 enables us to constrain the mass-loss rate to $\dot{M} \lesssim \rm 1.2\times10^{-5}\,M_{ \odot}\,yr^{-1}$. A tighter constrain on $\dot{M} \lesssim \rm 1\times10^{-5}\,M_{ \odot}\,yr^{-1}$ can be made by using the local {\it rms} of 18 $\mu$Jy\,beam$^{-1}$.
In comparison, for WR\,142, the non-detection agrees well with the flux densities predicted by the model. 
We highlight that mass-loss rates derived from thermal radio emission have been argued to be more accurate and robust than those from complex spectroscopic modelling approach. Estimates from radio emission are mostly based on observable quantities (see Eq.~\ref{equ:S_nu}) unlike the other methods which are model-dependent.
In fact, radio observations have been widely used to determine or constrain mass-loss rates in massive stars \citep[e.g.,][]{{Abbott1986},{Cappa2004},{DeBecker2019},{Benaglia2019},{Gallego2021},{Moutzouri2022}}; a comprehensive discussion on the advantages and limitations of different methods can be found in \citet{Puls2008}.

Besides, the scenario of NT radio emission associated with single (non-runaway) WR stars deserves to be discussed as well. On the one hand, local instabilities \citep[e.g.,][]{Lucy1980,White1985} or magnetic confinement \citep[e.g.,][]{Jardine2001} have been proposed to give rise to particle acceleration in single stellar winds. However, this has never been confirmed by the numerous radio observations of massive stars over the past decades. On the other hand, the termination shocks of stellar bubbles powered by WR stars are more likely to offer the required conditions to drive efficient particle acceleration. This is seen in the case of the WR bubble G2.4+1.4 around WR\,102, where the detection of synchrotron emission was reported by \citet{Prajapati2019} on a bubble of radius $\sim$2.5~pc. 
For WR\,114, there is no associated stellar bubble detected in the \textit{GLIMPSE}\footnote{Galactic Legacy Infrared Mid-plane Survey Extraordinaire \citep[\textit{GLIMPSE};][]{Benjamin2003}.} or \textit{WISE}\footnote{Wide-field Infrared Survey Explorer \citep[\textit{WISE};][]{Wright2010}.} images. The field of WR\,142 is comparatively complex. It is located at the edge of a large-scale superbubble/shell associated with Cyg~OB1 harbouring multiple WR and Of stars. 
\cite{Lozinskaya1991} discuss about an infrared shell of $\sim 3\degree$ but rule out the likelihood of its formation due to WR\,142 superwind.
The formation, size, and lifetime of these bubbles depend on several factors such as the progenitor evolutionary phase, wind parameters, mass ejected in previous evolutionary phases, the local ISM density, lifetime in the WR phase, the proper motion of the star, swept-up matter (ISM or stellar ejecta), and the dissipation rate of wind energy. Observationally, only a small fraction of known Galactic WR stars are seen to be associated with ring nebulae or stellar bubbles \citep[e.g.][]{Lozinskaya1991,Freyer2006}.  
In spite of being similar in spectral type and wind kinematics to WR\,102, the WR\,142 system does not exhibit any nearby associated stellar bubble. A quantitative comparison is quite difficult given that most of the above mentioned parameters are unknown for both stars. 
The non-detection of any wind-ISM interaction regions associated with WR\,142 motivates to further investigate the case of WR\,102 and the origin of its synchrotron emission.

\subsubsection{Colliding-wind binary (CWB) systems}
WR\,114 has been speculated to have a binary companion \citep{vanderHucht2001}. 
Similarly, the possibility of binary was also suggested for WR\,142, in which \citet{Sokal2010} suggested that a colliding-wind shock model --where the wind from the WR star shocks against an undetected companion or its wind-- could explain the high temperature indicated by the X-ray spectra.
Considering these inferences, in the following analysis we interpret the observational results assuming that both targets are PACWBs. 

In PACWBs, low frequency radio emission is most likely to be a combination of (i) free-free emission from the stellar wind of each binary component, and (ii) synchrotron radiation from the wind-collision region, where relativistic electrons are accelerated. 
The thermal radiation is expected to follow the free-free emission model discussed in Sect.~\ref{sec:results-singlestellarsys}. However, estimating the total flux density in these systems is not straightforward.

\citet{DeBecker2013} discussed the energy budget of stellar winds in PACWBs. In these systems, a fraction of the wind kinetic power is converted to the final radio emission after going through a series of energy conversion processes \citep[see Fig.~3 in][]{DeBecker2013}. Apart from the wind kinetic power and the efficiency of its conversion to particle acceleration giving rise to NT radiation, the observed flux density will also be influenced by FFA that depends on the orbital phase, orientation and wind opacity. 
Thus, the radio synchrotron emission is directly proportional to the wind kinetic power, where the proportionality is dependent on several variables that can vary significantly from one system to the other, and as a function of the orbital phase for a specific one. 

A radio synchrotron efficiency (RSE) parameter, that essentially gives the fraction of wind kinetic power converted to radio emission, is defined as \citep{DeBecker2013,DeBecker2017}
\begin{equation}
    \mathrm{RSE} = \frac{L_{\rm radio}}{P_{\rm kin}}.
\end{equation}
With the census of radio measurements and the information of $P_{\rm kin}$ from the PACWB catalog of \citet{DeBecker2013}, and based on a few selected objects, \citet{DeBecker2017} proposed empirical expressions to estimate the lower and upper boundary of RSE through a quadratic regression \citep[see Fig.~4 in][]{DeBecker2013}. 
These equations essentially outline the upper and lower limits of RSE as a function of $P_{\rm kin}$. Subsequently, these can be translated to the expected range of radio luminosities for a system with known $P_{\rm kin}$. The estimated RSE ranges from $10^{-9.7}$ to $10^{-7.6}$ and $10^{-9.9}$ to $10^{-7.8}$ for WR\,114 and WR\,142, respectively.
The radio luminosity ($L_\mathrm{radio}$) is estimated from the RSE limits, which can then be converted to flux density using the following equation:
\begin{equation}
    S_{\nu}^{\rm NT} = \frac{10^{26}\,(\alpha + 1)\, {\nu}^{\alpha}\,{L}\rm _{radio}}{4\pi D^2 \,( {\nu}^{(\alpha +1)}_2 - {\nu}^{(\alpha +1)}_1 )} \quad \mathrm{mJy}
\end{equation}
where $S_{\nu}^{\rm NT}$ 
is the flux density from NT emission at the observing frequency ($\nu$ in Hz), $\alpha$ is the spectral index, and $D$ is the distance to the source expressed in cm. Using the lower limit of radio luminosity, $\nu_1 = 0.1$~GHz, $\nu_2 = 50$~GHz, and adopting $\alpha = -0.5$ (canonical value for synchrotron radiation from relativistic electrons accelerated by DSA in high Mach number adiabatic shocks), we estimate the lower limit of the flux density at different frequencies $\nu$. The resulting SEDs including free-free and synchrotron emission are shown in Fig.~\ref{fig:flux_predict}.

For both sources, the (minimum) expected value is much higher than the 3$\sigma$ upper limits from our observations. For WR\,114, the expected total flux densities are 0.58 and 0.55~mJy at 735 and 1260~MHz, respectively. And in the case of WR\,142, at 735 and 1260~MHz, the flux density values are 0.50 and 0.41~mJy, respectively. Thus, if NT emission were present, it should have been detected with a high signal-to-noise ratio in our uGMRT observations. We note that this inference holds also if one assumes a spectral index $\alpha < -0.5$ in the calculation, as seen in some PACWBs \citep{DeBecker2013}. The non-detection in the lower frequency uGMRT bands, that are ideal for probing NT emission, implies an exceptionally low RSE (RSE $<\,10^{-10}$). This opens up a few possibilities that are briefly discussed below:
\begin{enumerate}
    \item The stars are not in binary systems with a massive companion with strong winds, leading to a lack of magneto-hydrodynamic shocks with the required properties for particle acceleration.
    \item The stars are in very wide binary systems (with period of decades). 
    In a CWB, the emission from the wind-collision region depends on the stellar separation. In systems with eccentric orbits, the (intrinsic) synchrotron emission peaks towards periastron, when the stellar separation is the least. It is possible that the two WR stars are in a very long-period orbit, with one of the binary components far from the periastron, leading to a very low synchrotron luminosity.
    \item The stars are in close binary systems.
    The wind-collision region would be within the high opacity zone of the radio photospheres of the binary components (see Sect.\,\ref{sec:photosphere}), so that radio emission would be drastically reduced due to FFA. From the estimated stellar wind photospheres, the orbital separation would be less than 70~AU and 20~AU for WR\,114 and WR\,142, respectively. 
    This is remarkably consistent with the analysis of X-ray observations of WR\,142 by \citet{Sokal2010}, who showed that the presence of a close (separation of $\sim$1~AU) B0V companion can explain the observed X-ray luminosity in a colliding-wind shocks scenario. 
    A similar scenario has also been inferred for WR\,133 \citep{DeBecker2019} and WR\,11 \citep{Benaglia2019}, where the high optical depth zone is much larger than the typical dimension of the full binary system.
\end{enumerate}

\section{Conclusions}
\label{sec:conclusion}
We carried out a low-frequency (735 and 1260~MHz) radio continuum study of two WR stars, WR\,114 and WR\,142, using uGMRT observations with high angular resolution and sensitivity. At both frequencies, no radio emission is detected from either of the WR stars. Based on the non-detection, we report on 3$\sigma$ upper limits to the radio flux densities at 735 and 1260~MHz (123 and 66~$\mu$Jy for WR\,114, and 111 and 96~$\mu$Jy for WR\,142, respectively). These limits are discussed in the context of two scenarios: stellar wind from a single star and a CWB system.

Considering the thermal emission from the stellar wind of a single star enables us to constrain the mass-loss rate of WR\,114 to $\dot{M}\,\lesssim \rm 10^{-5}\,M_{ \odot}\,yr^{-1}$, which is a factor of $\sim$3 lower than the model-based value from \citet{Sander2019}. 
In addition, our analysis does not reveal any synchrotron emission associated to a potential wind-ISM interaction. A comparison to the synchrotron-emitting WR bubble around WR\,102 points to the requirement to carefully consider the properties of the direct circumstellar environment to envisage significant particle acceleration at a termination shock.

If we rather assume both WR stars to be in CWB systems, the non-detection of NT emission in the lower-frequency uGMRT maps suggests either (a) a very wide binary system (with an orbital period of several decades) not close to periastron, or (b) a close binary system with strong FFA. Alternatively, the non-detection of synchrotron emission may also indicate the lack of a companion. The latter point deserves complementary studies at other wavebands to independently constrain the multiplicity of these two WR stars.

The investigation of particle acceleration from massive stars in various configurations, either single stars or in binary/multiple systems, definitely deserves to be pushed forward with more targets investigated at various radio frequencies, at various epochs, and keeping track of all potential shock physics scenarios likely to occur in these objects. Our results show that the availability of a quite high kinetic power is not sufficient to warrant the identification of clear indicators of particle acceleration in these objects. In particular, for isolated WR stars, the efficiency of converting wind kinetic power into particle acceleration can be rather low under unfavourable conditions.

\section*{Acknowledgements}

We thank the referee for valuable suggestions that helped us to improve the manuscript. This work was carried out in the framework of the PANTERA-Stars ({\url{https://www.astro.uliege.be/~debecker/pantera/}}) initiative. We thank the staff of the GMRT that made these observations possible. GMRT is run by the National Centre for Radio Astrophysics of the Tata Institute of Fundamental Research. 

\section*{Data Availability}

The raw data is publicly available in the GMRT webpage (\url{https://naps.ncra.tifr.res.in/goa/data/search}).
The calibrated data used in this article will be shared upon reasonable request to the corresponding author.



\bibliographystyle{mnras}
\bibliography{reference} 








\bsp	
\label{lastpage}
\end{document}